\documentstyle[aps,multicol,prl,epsfig,floats]{revtex}
\begin{document}

\twocolumn[\hsize\textwidth\columnwidth\hsize\csname@twocolumnfalse%
\endcsname

\title{Critical behavior of the frustrated antiferromagnetic six-state
       clock model on a triangular lattice}

\author{J. D. Noh and H. Rieger}
\address{Theoretische Physik, Universit\"at des Saarlandes,
66041 Saarbr\"{u}cken, Germany}

\author{M. Enderle, and K. Knorr}
\address{Technische Physik, Universit\"at des Saarlandes,
66041 Saarbr\"{u}cken, Germany}

\maketitle

\begin{abstract}
  We study the anti-ferromagnetic six-state clock model with nearest
  neighbor interactions on a triangular lattice with extensive
  Monte-Carlo simulations. We find clear indications of two phase
  transitions at two different temperatures: Below $T_I$ a chirality
  order sets in and by a thorough finite size scaling analysis of the
  specific heat and the chirality correlation length we show that this
  transition is in the Ising universality class (with a non-vanishing
  chirality order parameter below $T_I$). At $T_{KT}(<T_I)$ the
  spin-spin correlation length as well as the spin susceptibility
  diverges according to a Kosterlitz-Thouless~(KT) form and spin
  correlations decay algebraically below $T_{KT}$. We compare our
  results to recent x-ray diffraction experiments on the orientational
  ordering of CF$_3$Br monolayers physisorbed on graphite. We argue that
  the six-state clock model describes the universal feature of the
  phase transition in the experimental system and that 
  the orientational ordering belongs to the KT universality class.
\end{abstract}

\pacs{75.10.Hk, 64.60.Cn, 75.40.Mg}

]

\section{Introduction}

The study of frustrated two-dimensional spin models is motivated by
their relevance for phase transitions in a wide range of physical
systems. For instance, the fully frustrated $XY$~(FF$XY$) model
describes an array of Josephson junctions under 
an external magnetic field~\cite{Teitel83,lee}.
Besides its experimental relevance the FF$XY$ model has been 
the focus of a number of theoretical works since a delicate question arises 
here regarding the critical behavior. 
The model has a continuous $U(1)$ symmetry {\it and} a discrete Ising,
chiral or $Z_2$ symmetry that can be broken at low temperatures 
through a Kosterlitz-Thouless~(KT) type
transition and an Ising-like transition, respectively.

Despite a lot of efforts~[3--22], there is no consensus on the nature of 
the phase transitions and it is not clear whether the two transitions 
happen at two different temperatures or at a single one.
Renormalization group~(RG) studies on the FF$XY$ model drew the conclusion
that the transitions occur at the same 
temperature~\cite{Choi85&Yosefin85}.
Monte Carlo simulation studies on generalized FF$XY$ models supported
the single transition picture for the FF$XY$ model~\cite{Berge86&Eikmans89}.
Interestingly, it has been reported that the $Z_2$ symmetry breaking 
transition may not belong to the Ising universality class. 
Monte Carlo studies of
the FF$XY$ models on a triangular and a square lattice yielded the
correlation length exponent $\nu= 0.83(4)$~(triangular lattice)
and $\nu = 0.85(3)$~(square lattice), which is inconsistent with
the Ising value $\nu=1$~\cite{JLee91}. The specific heat appeared
to follow a power-law scaling rather than the logarithmic
scaling as one would expect for the Ising universality class. 
A non-Ising scaling behavior is also observed in the studies of 
the FF$XY$ model via a Monte Carlo
simulation~\cite{R-Santiago92} and Monte Carlo transfer matrix
calculations~\cite{Granato93,Knops94}. The same non-Ising critical
behavior is also observed in the coupled $XY$-Ising
model~\cite{JLee91,Granato91,Nightingale95}. 

On the other hand, numerical evidence also in favor of two transitions
at two different temperatures has been collected.
Monte Carlo simulation studies of the frustrated Coulomb gas system, which
is supposed to be in the same universality class as the FF$XY$ model,
showed that the KT type transition temperature $T_{KT}$ and the Ising-like
transition temperature $T_{I}$ are different with
$T_{KT}<T_I$~\cite{Grest89,JRLee94}. Two transitions were also found
in the FF$XY$ model on a square lattice~\cite{SLee94} and
on a triangular lattice~\cite{Miyashita84,Xu96,SLee98} using 
Monte Carlo simulations.
A careful analysis of the RG flow of the FF$XY$ model also led to a
conclusion of the double transition scenario~\cite{Jeon97}. 
It is the general belief that the transitions at $T_{KT}$ and at 
$T_I$ belong to the KT universality class~\cite{KT73} and to the
Ising universality class, respectively, if the transitions occur at
different temperatures. However, the critical exponents associated with
the $Z_2$ symmetry breaking, which are found by Monte Carlo simulations,
turn out to be different from those of the Ising
universality~\cite{JRLee94,SLee94,SLee98}. They are rather
close to those obtained in the $XY$-Ising model~\cite{JLee91}.
Although there is an argument that the observed non-Ising exponents
are due to a screening effect hindering the asymptotic scaling
behavior~\cite{Olsson95}, the controversy on the nature of the phase 
transition remains unsettled~\cite{Comment,Boubcheur98,Loison00}.

Here we present a thorough numerical study of a related model, the
fully frustrated anti-ferromagnetic six-state clock model on a
triangular lattice. We also report on an orientational ordering
transition in a two-dimensional experimental system~\cite{Fassbender02} 
that, as we argue,
is in the same universality class as the model we study numerically.
The experimental system actually motivates (besides numerical
simplicity) our restriction to the six states of the spins 
rather than the continuum $XY$ spins (and thus to a 
six-fold clock~($C_6$) symmetry rather than the $U(1)$ symmetry). 
In the unfrustrated (ferromagnetic) case 
it has been shown, however, that the KT-behavior is stable with respect 
to a crystal field of six-fold 
symmetry~\cite{Jose_etal,Itakura01,Challa_Landau86}. 
Hence we expect that our model displays the same critical behavior
as the FF$XY$ model.

The paper is organized as follows: In Sec.~\ref{sec:2} 
we define the model and identify symmetry in the ground state.
The corresponding order parameters are also defined.
In Sec.~\ref{sec:3} we explain briefly the Monte-Carlo
procedure we used and determine the transition temperatures for the
two phase transitions. Section~\ref{sec:4} is devoted to the 
classification of the universal properties at the two transitions 
we find. In Sec.~\ref{sec:5} we present experimental x-ray diffraction 
results on the orientational ordering of CF$_3$Br monolayers physisorbed 
on graphite which we expect to be described by the KT-transition 
occurring in the model we studied numerically in the preceding sections. 
Section~\ref{sec:6}
concludes the paper with a summary of our results.

\section{Model}\label{sec:2}

\begin{figure}
\centerline{\epsfig{file=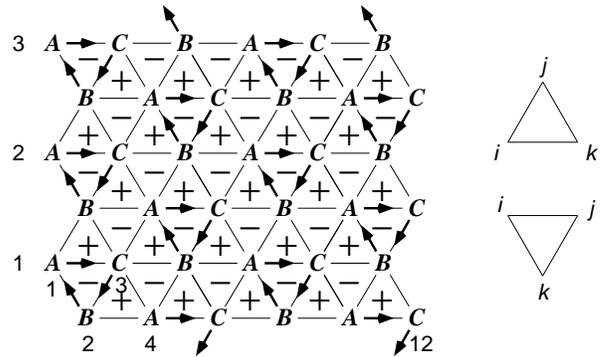,width=0.9\columnwidth}}
\caption{A triangular lattice of size $12\times 3$. 
The arrows represent the spin state in one of the ground states, ${\cal C}_0$.
The lattice sites are labeled by the sublattice indices $A$, $B$, and $C$. 
Also shown are elementary up- and down- triangles denoted by
$\bigtriangleup_i$ and $\bigtriangledown_i$, respectively.
The $\pm$ signs denote the chirality of the triangles.}\label{tri-lattice}
\end{figure}

We investigate the phase transitions of the anti-ferromagnetic 
six-state clock model on a two-dimensional (2D) $N = L_x\times L_y$ 
triangular lattice~(see Fig.~\ref{tri-lattice}). 
The six-state clock spin ${\bf S}$ is a planar spin pointing
toward discrete six directions; ${\bf S} = (\cos\theta,\sin\theta)$ with
\begin{equation}
\theta = \frac{2\pi n}{6} \quad (n = 0,1,\ldots,5) \ .
\end{equation}
The interaction is given by the Hamiltonian
\begin{equation}
{\cal H} = 2 J \sum_{\langle i,j\rangle} \cos(\theta_i - \theta_j) \ ,
\label{ham}
\end{equation}
where the sum is over all nearest neighbor site pairs $\langle i,j\rangle$ 
and $J>0$ is the antiferromagnetic coupling strength.
The overall factor $2$ is introduced for a computational convenience.

The antiferromagnetic interaction on a triangular lattice induces a
frustration. As a result the spins on each triangles should make an angle of
$\pm 120^\circ$ with one another in the ground state. 
There are 12-fold degenerate ground states with three-sublattice structure. 
According to their chirality, we can categorize the ground states 
into ${\cal C}_n$ and ${\cal A}_n$~($n=0,1,\ldots,5$). 
In each ground state the spin configurations of the three sublattices $A$,
$B$, $C$~(Fig.~\ref{tri-lattice}) are given by
\begin{equation}
\theta_{i\in A} = \frac{2\pi n}{6} \ ,
\theta_{i\in B} = \frac{2\pi(n\pm2)}{6}\  ,
\theta_{i\in C} = \frac{2\pi (n\pm4)}{6}
\end{equation}
for ${\cal C}_n$~(upper sign) and ${\cal A}_n$~(bottom sign).
In other words, in the ${\cal C}$(${\cal A}$) type ground state, 
the spin angles increase by $\frac{2\pi}{3}$ along the
up-triangles in the (anti-)~clockwise direction, and vice versa for 
down-triangles.

There are two different symmetries between them; those in the same class of
${\cal C}$ or ${\cal A}$ are related by the rotation
$\theta_i \rightarrow \theta_i + \frac{2\pi m}{6}~(m = 0,\ldots,5)$
and those in the others by the reflection
$\theta_i \rightarrow -\theta_i$.
Therefore the model possesses the $C_6~(\mbox{six-fold clock})$ symmetry
and the $Z_2~(\mbox{Ising})$ symmetry. Note that the FF$XY$ model on a
triangular lattice has a $U(1)$ and a $Z_2$ symmetry~\cite{lee}.
Since the clock spin can have only six states, while the $XY$ spin 
is a continuous one, the continuous $U(1)$ symmetry of the FF$XY$ model
is reduced to the discrete $C_6$ symmetry.
The nature of the $Z_2$ symmetry in each model is identical.

From the symmetry consideration, we expect that there
are two types of phase transitions associated with the spontaneous 
breaking of the $C_6$ symmetry and the $Z_2$ symmetry.
The chirality at each elementary triangle is defined as
$$
h_{\bigtriangleup_i,\bigtriangledown_i} = \frac{2}{3\sqrt{3}} 
\left[\sin(\theta_j-\theta_i)\!+\!
       \sin(\theta_k-\theta_j) \!+ \!\sin(\theta_i-\theta_k)\right] \ ,
$$
where $\bigtriangleup_i$ and $\bigtriangledown_i$ are up- and down-triangles
as depicted in Fig.~\ref{tri-lattice}.
Then the ground state ${\cal C}_n~({\cal A}_n)$ has a checker-board pattern
of the chirality with $h_{\bigtriangleup} = +1~(-1)$ and 
$h_\bigtriangledown=-1~(+1)$. The staggered chirality
\begin{equation}
h = \frac{1}{2N} \sum_{i} \big( h_{\bigtriangleup_i} - h_{\bigtriangledown_i} 
\big)
\end{equation}
plays a role of the order parameter for the $Z_2$ symmetry breaking
transition.
The order parameter for the $C_6$ symmetry breaking transition 
is the sublattice magnetization
\begin{equation}
m_A = \frac{1}{N} \sum_{i\in A} \exp(i\theta_i) \ ,
\end{equation}
where the sum runs only over the sites in the $A$ sublattice. $m_B$ and $m_C$
are defined analogously.

\section{Transition temperatures}\label{sec:3}

We performed Monte Carlo~(MC) simulations on finite $N=L\times (L/2)$ 
lattices with a sublattice updating scheme;
one of the three sublattices is selected randomly and then all spins in the
chosen sublattice are flipped according to the Metropolis rule.
One Monte Carlo step corresponds to three sublattice updates.
Various observables are measured during the MC runs, such as the
energy, the chirality, and the sublattice magnetization, from which we
can measure the averaged quantities and their fluctuations.
In some cases, histograms are constructed from particularly long MC
runs to obtain the observables as continuous functions of the temperature.

\begin{figure}[t]
\centerline{\epsfig{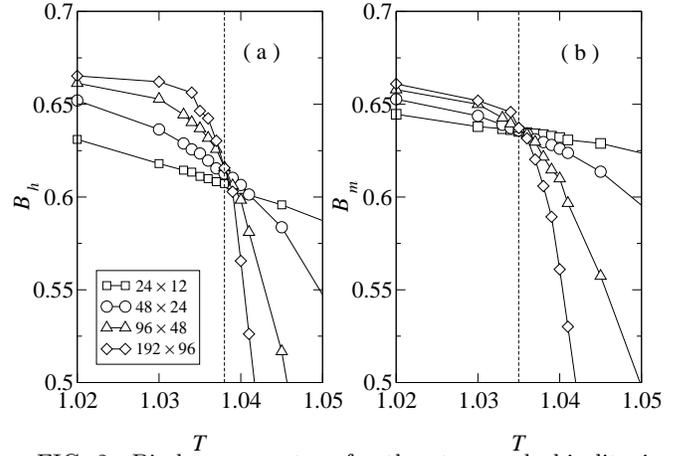}}
\caption{Binder parameters for the staggered chirality in (a) and the
magnetization in (b). The estimated transition temperatures are marked by
broken lines.}
\label{Bh_Bm}
\end{figure}

The transition temperature is determined from the Binder parameter
\begin{equation}
B_h = 1 - \frac{\langle h^4\rangle}{3 \langle h^2\rangle^2} \ \ \mbox{and} \ \
B_m = 1 - \frac{\langle m_A^4\rangle}{3 \langle m_A^2 \rangle^2} 
\end{equation}
for the chirality and the magnetization, respectively.
Here the angle bracket denotes a thermal average, which can be done by 
a time average over the MC runs.
The Binder parameter is scale-independent at the critical points and 
approaches $\frac{2}{3}$~(0) in the ordered~(disordered) phase
as $L$ becomes larger. Hence the
order-disorder transition temperature $T_I$ related to the $Z_2$ 
symmetry breaking is obtained from the crossing point in the plot of 
$B_h$ versus $T$ at different system sizes.
From Fig.~\ref{Bh_Bm} (a), we estimate that 
\begin{equation}\label{T_I}
T_I = 1.038 \pm 0.0005 \ .
\end{equation}
$B_m$ also displays the crossing behavior~(Fig.~\ref{Bh_Bm} (b)), from which
we estimate that
\begin{equation}\label{T_KT}
T_{KT} = 1.035  \pm 0.0005 
\end{equation}
for the $C_6$ symmetry breaking transition.

\begin{figure}[t]
\centerline{\epsfig{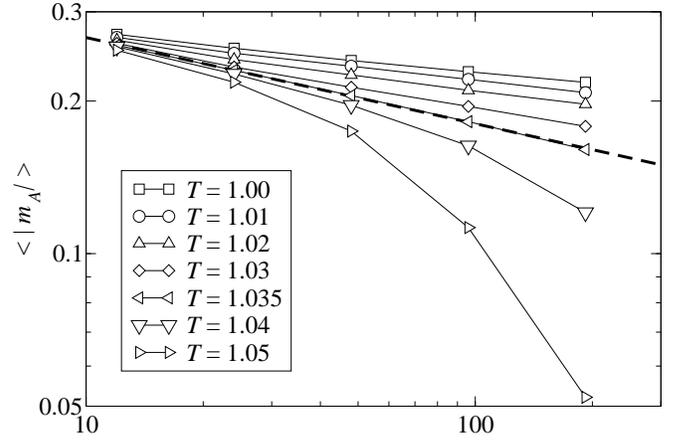}}
\caption{Sublattice magnetization $\langle | m_A | \rangle$ near the
critical temperature $T_{KT}$. At temperatures $T\leq T_{KT}$ it shows
a power law behavior. The broken line has a slope $-0.17$.}
\label{mA}
\end{figure}

The Binder parameters show different size dependence at low
temperatures. For the staggered chirality $B_h$ converges to $2/3$
rapidly as $L$ increases. However, $B_m$ appears to converge
to values less than $2/3$ at $T<T_{KT}$. It indicates that there is a
quasi-long-range order in spins at $T<T_{KT}$.
We confirm it from finite-size-scaling~(FSS) behaviors of the
magnetization. The sublattice magnetization shows the power law scaling
behavior, $\langle | m_A | \rangle \sim L^{-x}$, at $T\leq T_{KT}$  
with temperature dependent exponent $x$~(see Fig.~\ref{mA}).

\begin{figure}[t]
\centerline{\epsfig{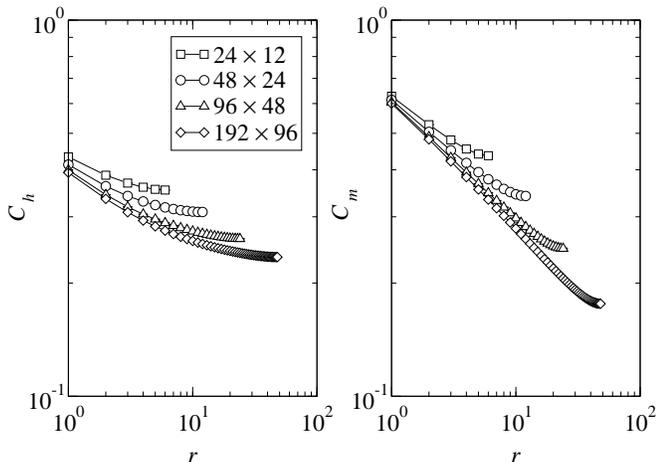}}
\caption{Correlation functions of the staggered chirality, $C_h$,
and the magnetization, $C_m$, at temperature $T=1.037$.}
\label{corr_1.037}
\end{figure}

The two transition temperatures lie very close to each other. Nevertheless
the accuracy of the data is sufficiently high that we can exclude a 
possibility $T_{I} = T_{KT}$. Figure~\ref{corr_1.037} demonstrates
this via the correlation functions. 
We measure the correlations between the chirality of up-triangles 
and the magnetization of spins in the $A$ sublattice displaced by 
a distance $r$ in the vertical direction:
\begin{equation}\label{c_h}
C_h (r) = \left \langle \frac{3}{N} \sum_{i\in A} h_{\triangle_i}
h_{\triangle_{i+r}} \right\rangle  \ ,
\end{equation}
\begin{equation}\label{c_m}
C_m (r) = \left \langle \frac{3}{N} \sum_{i\in A} \cos( \theta_i -
\theta_{i+r}) \right\rangle \ .
\end{equation}
Here, $i+r$ denotes a site displaced vertically by a distance 
$r$ from $i$, and $\triangle_i$ denotes an up-triangle whose left
corner is $i$~(see Fig.~\ref{tri-lattice}). 
In Fig.~\ref{corr_1.037}, we plot both correlation functions in the log-log 
scale at an intermediate temperature $T = 1.037$. 
Clearly one can see an upward curvature in the plot of $C_h(r)$ for
$r\ll L_y/2$,
which implies that the chirality order has already set in. On the other hand,
there is a downward curvature in the plot of $C_m(r)$ for $r\ll L_y/2$ indicating that
the magnetic order has not set in yet. Therefore we conclude that
$T_{KT} < 1.037 < T_I$.

\section{Universality class}\label{sec:4}
With $T_{KT}\neq T_{I}$, one expects that the both symmetry breakings
take place independently. 
Then, the $Z_2$ symmetry breaking transition should be in the 
Ising universality class.
The $C_6$ symmetry is equivalent to that of the $XY$ model perturbed by the
$(p=6)$-fold anisotropy field. The unperturbed $XY$ model displays
a KT transition which separates a disordered high-temperature
phase and a quasi-long range ordered low-temperature phase. A
renormalization group~(RG) study shows that an anisotropy field
with $p>4$ does not change the KT nature of the transition~\cite{Jose_etal}.
The same is true for the extreme case of the ferromagnetic six-state clock
model~\cite{Challa_Landau86}. So the $C_6$ symmetry breaking phase transition 
at $T=T_{KT}$ is expected to belong to the KT universality class.

The interplay of the Ising type and the KT type ordering has long been
studied in the context of the fully frustrated $XY$ models on a square
or a triangular lattice. 
Frequently, it has been 
claimed~\cite{Choi85&Yosefin85,Berge86&Eikmans89,JLee91,R-Santiago92,Granato93,Knops94}
that the Ising-like transition 
and the KT-type transition occur at the same temperature.
On the other hand, more recent studies have reported the transitions to occur
at two different
temperatures~\cite{Grest89,JRLee94,SLee94,Miyashita84,Xu96,SLee98,Jeon97,Olsson95,Comment,Loison00}. 
Surprisingly, some of recent high-precision Monte Carlo simulation studies 
suggested that the $Z_2$ symmetry breaking transition
does not belong to the Ising universality class in spite of the double
transition~\cite{SLee94,SLee98,Loison00}. 
They estimated that $\nu\simeq 0.8$ for the correlation length
exponent and $\alpha/\nu\simeq 0.46$ with the specific exponent $\alpha$, 
which are incompatible with the Ising values of $\nu=1$ and $\alpha=0$.
The critical exponents are consistent with the values observed along the
single transition line of the coupled $XY$-Ising
model~\cite{JLee91,Granato91,Nightingale95}. Olsson~\cite{Olsson95} argued
that the measured non-Ising critical exponents are artifacts of unusual
finite size effects originated from non-critical spin-wave fluctuations.
However, apparently there is no general agreement yet on this 
subject~\cite{Comment}.

In our model, we observe the phase transitions occurring at two 
{\em different} temperatures. However, the controversy existing in the
FF$XY$ model tells us that it does not guarantee necessarily that the 
individual transitions will belong to the KT universality class and the Ising
universality class, respectively. Therefore we perform a thorough 
finite-size-scaling~(FSS) analysis to understand the nature of the 
phase transition.
We believe that the antiferromagnetic six-state clock model can be
used to resolve the existing controversy for the FF$XY$ model. 
It has the proper symmetry properties as discussed
previously. And, from a practical point of view, larger system sizes
are available since it is a discrete model, with which one can
reduce the finite size effects.

\subsection{$Z_2$ symmetry breaking}

We investigate the nature of the $Z_2$ symmetry breaking
transition by exploring the FSS property of the specific heat, defined by
\begin{equation}
c = N \left[ \langle e^2 \rangle - \langle e\rangle^2 \right]
\end{equation}
with $e$ the energy per site. For a finite system ($N = L\times
L/2$) the specific heat has a peak $c_* \sim
L^{\alpha/\nu}$ near the critical temperature, diverging with system size
$L\rightarrow\infty$. 
Since the specific heat does not diverge at a KT type transition, the
divergence is due to the $Z_2$ symmetry breaking phase
transition. The Ising universality class has $\alpha=0$, and the
specific heat shows a logarithmic divergence, $c_* \sim \ln L$.

\begin{figure}
\centerline{\epsfig{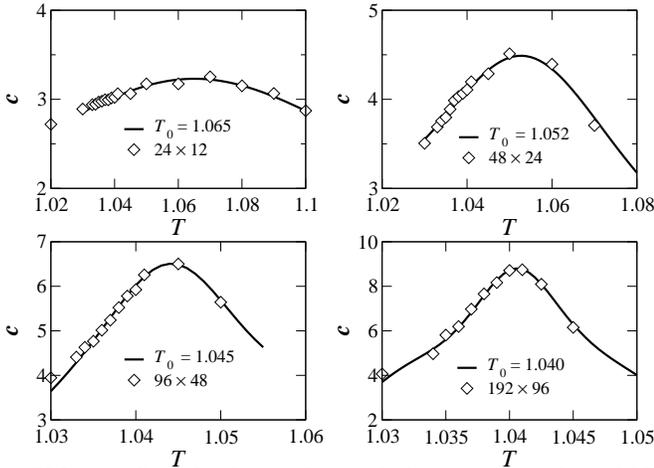}}
\caption{Specific heat obtained from independent MC runs~(symbol) and
the histogram method~(solid lines). $T_0$ denotes the temperature 
at which the histogram is obtained.}
\label{cv_max}
\end{figure}

The specific heat is measured accurately using the standard Monte
Carlo~(MC) method combined with a {\em single histogram
method}~\cite{Ferrenberg}.
First we measure the specific heat from the standard MC simulations 
at discrete temperature grid, which leads to a temperature $T_0$ at which the
specific heat is maximum approximately. Then the histogram $\Omega_0(\{{\bf
S}\})$, frequency of a spin state $\{{\bf S}\}$, is constructed
from MC runs at the temperature $T_0$ over a time interval $\Delta
t$. It enables us to obtain the average value of any observable
$Q$ as a continuous function of temperature $T$ near $T_0$;
\begin{equation}
\langle Q \rangle = \frac{\sum_{\{\bf S\}} Q(\{{\bf S}\})\
\Omega_0 (\{{\bf S}\})\ e^{-(T^{-1}-T_0^{-1}) E(\{{\bf
S}\})}}{\sum_{\{\bf S\}} \Omega_0(\{{\bf S}\})\
e^{-(T^{-1}-T_0^{-1}) E(\{{\bf S}\})}} \ .
\end{equation}
It is important to take $\Delta t$ as large as possible 
to have good statistics. At $192\times 96$ lattice, for
example, we take $\Delta t = 5 \times 10^7$ which is $\sim 2000$
times of the relaxation time of energy-energy auto-correlation.
The maximum system size we can simulate is $384\times 192$. In
that case we take $\Delta t = 10^8$ which is $\sim 1200$ times of
the relaxation time of energy-energy auto-correlation.
Figure~\ref{cv_max} shows the specific heat obtained from
independent MC runs and from the histogram method. Their agreement
near the peak is excellent, although they begin to deviate slightly
away from the peak. 
With this technique we can estimate $c_*$ very accurately. 
A statistical error
for $c_*$ is estimated in the following way; we construct ten
sub-histograms and as many values of $c_*$ from each of them. The
error bar is taken as the half distance between the maximum and
minimum values among them.

\begin{figure}
\centerline{\epsfig{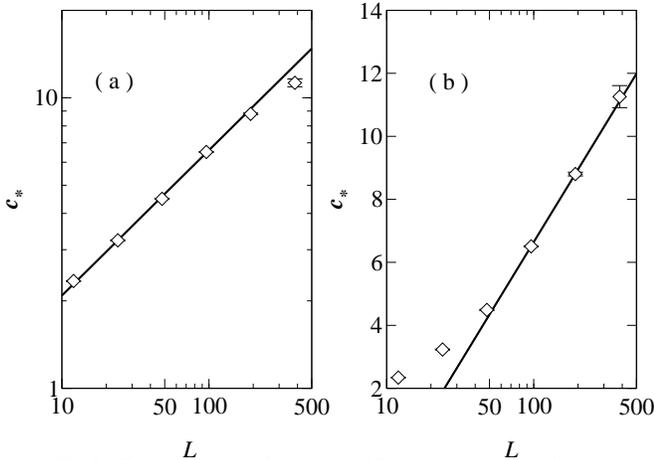}}
\caption{Peak values of the specific heat versus $L$ in log-log scale 
(a) and semi-log
scale (b). In (a) the solid line has a slope $0.5$.}
\label{cv_star}
\end{figure}

The result is shown in Fig.~\ref{cv_star}. Interestingly,
there is a crossover from a power-law scaling behavior $c_* \sim L^{0.5}$
(solid line in Fig.~\ref{cv_star}(a)) at $L \leq L_c$ to a
logarithmic scaling behavior $c_* \sim \ln L$ (solid line in
Fig.~\ref{cv_star}(b)) at $L \geq L_c$ with $L_c \simeq 96$. 
It indicates that the specific heat has a logarithmic scaling behavior 
asymptotically, which is consistent with the Ising universality class.

\begin{figure}
\centerline{\epsfig{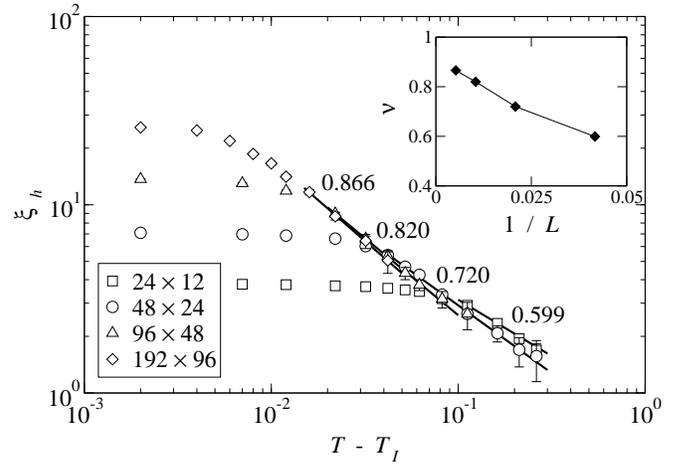}}
\caption{Chirality correlation length. The values of the exponent $\nu$
obtained from the fit at each lattice size are shown in the main panel and
plotted in the inset.}
\label{xi_h}
\end{figure}

We also study the scaling of the chirality correlation length. 
We measured the correlation length using
$\xi_h = \sqrt{\sum_r r^2 C_h(r) / \sum_r C_h(r)}$ with the correlation 
function defined in Eq.~(\ref{c_h}). It is plotted in Fig.~\ref{xi_h}. 
We find a strong dependence of $\xi_h$ on the system size. 
In addition to the saturation of $\xi_h$ to $O(L)$ near and below 
the critical point, the dependence of the correlation length on $L$ 
in the regime $\xi_h \ll L$ is visible through a varying 
slope (cf., Fig.~\ref{xi_h}). 
As a consequence, a fitting to a form $\xi_h =
a(T-T_I)^{-\nu}$ at small system size will lead to a lower value of
$\nu$ than its asymptotic value. The fitted values of $\nu$ at
each lattice size are plotted in the inset of Fig.~\ref{xi_h}.
It increases as the system size and approaches 
the Ising value $\nu \simeq 1.0$ in the infinite size limit. 

From the analysis of the specific heat and the chirality correlation length, 
we conclude that the $Z_2$ symmetry breaking transition indeed belongs to
the Ising universality class.
Note that the {\em effective} specific heat exponent $\alpha/\nu \simeq 0.5$ 
and the correlation length exponent $\nu\simeq 0.8$ appearing at small length scale 
are comparable with the results of previous
MC works claiming the non-Ising nature in the frustrated $XY$
models~\cite{JLee91,R-Santiago92,Granato91,Nightingale95,SLee94,SLee98,Boubcheur98,Loison00}. 
Our results are fully consistent with Olsson's argument claiming
that the non-Ising exponents reported by others are explained by
a failure of finite-size-scaling at small length scale due to the 
screening length associated with the nearby KT transition~\cite{Olsson95}.
The strong finite size effect is overcome in this work since we could
study large systems.
We expect that one could observe the same crossover to the asymptotic 
Ising type scaling behavior in larger scale simulations in the frustrated 
$XY$ systems.

\subsection{$C_6$ symmetry breaking}

\begin{figure}
\centerline{\epsfig{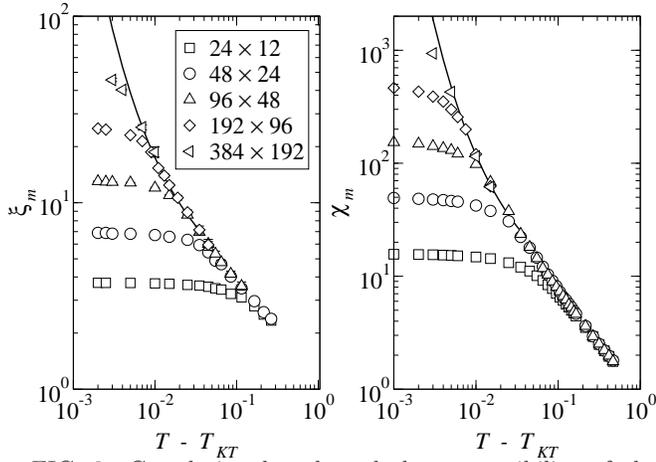}}
\caption{Correlation length and the susceptibility of the magnetization.
The solid lines are the fits to the KT form of $\xi_m = 2.54 e^{0.193
(T-T_{KT})^{-1/2}}$ and $\chi_m = 4.20 e^{0.333 (T-T_{KT})^{-1/2}}$.}
\label{KTscaling}
\end{figure}

The KT transition is characterized by the essential singularity of
the correlation length and the susceptibility approaching the
critical temperature. We measure the correlation length $\xi_m$
as
\begin{equation}\label{xi-from-C}
\xi_m = \sqrt{\frac{ \sum_r r^2 C_m(r)}{\sum_r C_m(r)}}
\end{equation}
from the magnetic correlation function $C_m$ in Eq.~(\ref{c_m}).
One can also estimate the correlation length by fitting $C_m(r)$
to a form with $e^{-r/\xi_m}$. We tried with various function
forms but found that Eq.~(\ref{xi-from-C}) yields the estimate most stable
against statistical errors. The magnetic susceptibility is
measured from the fluctuation of the magnetization, $\chi_m = N
\langle m_A^2 \rangle$. Figure~\ref{KTscaling} shows the plot of
$\xi_m$ and $\chi_m$ versus $T-T_{KT}$ in log-log scale. Upward
curvature in both plots indicates a divergence stronger than 
algebraic. We fit those quantities with the KT scaling
form, $\xi_m = a e^{b (T-T_{KT})^{-y}}$ and $\chi_m = a' e^{b'
(T-T_{KT})^{-y}}$ with $y=\frac{1}{2}$ fixed, whose results are
drawn with solid lines in Fig.~\ref{KTscaling}. It shows that the
KT scaling behaviors are emerging only after $L\geq 96$. If $y$
is also taken as a fitting parameter, one obtains $y<0.5$ whose
value depends on the fitting range; it increases as approaching
the critical temperature. 

In the $XY$ phase ($T<T_{KT}$) the spins have 
the quasi-long range order; the magnetization scales 
algebraically as $ \langle |m_A|\rangle
\sim L^{-x}$ with temperature dependent exponent $x$~(see
Fig.~\ref{mA}). The RG theory predicts that $x=1/8$ at the KT
transition point~\cite{Jose_etal}. 
Our numerical data show that $x\simeq 0.17$ at $T = T_{KT}$. 
We think that the discrepancy is originated from a sensitive dependence of
$x$ on $T$.
In summary, in spite of a quantitative disagreement, the $C_6$
symmetry breaking phase transition is in qualitative agreement
with the KT universality class; we observe the essential singularity 
in the correlation length and the susceptibility and quasi-long range
order in the low temperature phase.

\section{Experimental realization: CF$_3$Br on graphite}\label{sec:5}

In this section we describe a possible experimental realization of the
theoretical model we investigated above. In a recent work we reported
results on x-ray powder diffraction study on a monolayer of halomethane
CF$_3$Br adsorbed on exfoliated graphite \cite{Fassbender02}. CF$_3$Br is a
prolate molecule and has a dipole moment of about 0.5~D. The coverage
$\rho$, temperature $T$ phase diagram is rather complex
\cite{maus,knorr1}.  In \cite{Fassbender02} we concentrated on a coverage which
is representative of the extended monolayer regime in which the
monolayer lattice is commensurate with the graphite lattice. This
yields a $2\times 2$ triangular lattice arrangement of the CF$_3$Br molecules
below a temperature of 105~K \cite{knorr2}, which is the melting
temperature of the commensurate layer. The inter-molecular distance is
$a=4.92\;$\AA. Note that the lateral size of
the graphite crystallites is only around 180~\AA, which confines any
spatial correlation length to this value.

An isolated CF$_3$Br would prefer to lie flat on the substrate, but
the $2\times 2$ mesh is too tight to accommodate the molecules in
this orientation. Therefore the individual CF$_3$Br molecules stand on
the substrate, presumably with the F$_3$ tripod down, with a maximum
tilt angles of the molecular axis up to 30${}^\circ$ with respect to the
substrate normal due to steric repulsion. A tilt leads
to a non-zero in-plane component of the dipole moment. We
regard this component as planar pseudospin 
${\bf S}_i=(\cos\theta_i,\sin\theta_i)$ with the azimuthal angle $\theta_i$ 
of the molecule.  In this sense the $2\times 2$ state is disordered 
with a zero time average of every ${\bf S}_i$, and is
stabilized at higher temperatures by a libration and/or a precession of the
molecular axis about the substrate normal.

As the temperature is decreased additional features develop in the
diffraction pattern which finally, below 40~K, can be identified
\cite{Fassbender02} as Bragg peaks (with a finite width of around
(180~\AA)${}^{-1}$ due to the lateral size of the crystallites) indicating
an orientational order in the dipole moments identical to the one
depicted in Fig.~\ref{tri-lattice}. The temperature dependence of the
correlation length $\xi$, determined from the intrinsic width of this
peak, can be fitted with the KT-expression
\begin{equation}
\xi=A\exp(B(T/T_{KT}-1)^{-1/2})
\label{KT-form}
\end{equation}
to the data for $T>40\;$K~(see Fig.~3 of Ref.~\cite{Fassbender02}). 
The fit parameters are $A=9\pm2\;$\AA,
$B=1.5\pm0.4$, $T_{KT}=30\pm3\;$K. Note that the value of A is reasonably
close to the lattice parameter of the 2D mesh.  Thus $\xi$ is expected to 
diverge at a KT-critical temperature $T_{KT}$ of about 30 K,
but the growth of the correlated regions is interrupted when $\xi$ reaches
the size of the graphite crystallites. This happens at about 40 K.

We think that the model (\ref{ham}) is a good description of the
orientational ordering process described in this physical system: Clearly
the pseudospin correlations are bound to a plane, thus the system is
2D with respect to the relevant degrees of freedom at the phase
transition. Moreover, as mentioned before, below 105 K the CF$_3$Br
molecules are arranged in a triangular lattice. The pseudospin
representing the CF$_3$Br dipole moment is presumably not a strictly
isotropic planar rotator but experiences a crystal field from the graphite
substrate which breaks the continuous azimuthal symmetry into the six fold 
symmetry of the monolayer. 
This is reflected by the six-state clock variables of model
(\ref{ham}). The ordered structure of the CF$_3$Br dipoles is
antiferroelectric, which is taken into account by the antiferromagnetic
couplings between the pseudospins in (\ref{ham}). Finally, the
character of the relevant orientation dependent interactions of
CF$_3$Br is not known, but a comparison of monolayers of several polar
methane derivatives \cite{knorr1} shows that fully halogenated
molecules including CF$_3$Br with small dipole moments around 0.5~D
have structures different from partially halogenated molecules such as
CH$_3$Cl with strong dipole moments around 1.7~D. This suggests that
for CF$_3$Br the short range anisotropic part of the intermolecular
van der Waals force and hard-core repulsion are more important than
the medium range dipole-dipole interaction. Thus the interactions can
be assumed to be short ranged. Thus one expects that model (\ref{ham})
and the physical system discussed here are in the same universality
class.

\begin{figure}[t]
  \centerline{\epsfig{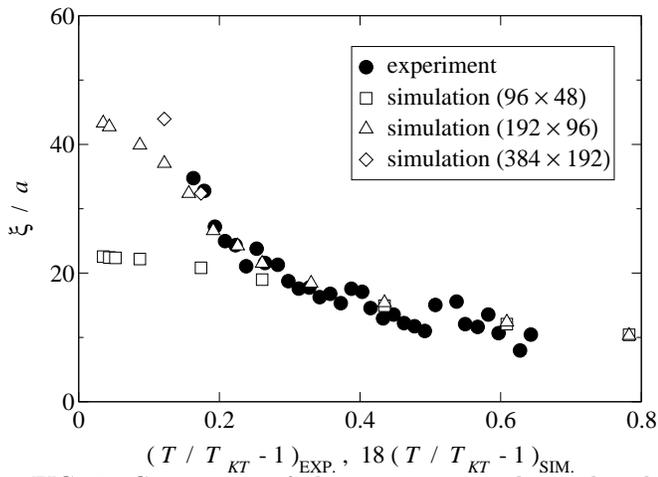}}
\caption{Comparison of the magnetic correlation length (scaled by the
  lattice constant) from the experimental data 
  (from Fig.~3 of Ref.~\protect\cite{Fassbender02}) and the Monte Carlo 
  simulations (from Fig.~\ref{KTscaling}).
  The reduced temperature $(T/T_{KT}-1)$ for the
  simulation data are rescaled by a factor of 18 in order to achieve
  an acceptable data collapse for linear lattice sizes $L$ between 96
  and 192.}
\label{exp-sim}
\end{figure}
In Fig.~\ref{exp-sim} a comparison of the magnetic correlation length
from the experiment~\cite{Fassbender02} and 
the simulation (from Fig.~\ref{KTscaling}) is shown. Since the
factor $B$ in the KT-form (\ref{KT-form}) of the correlation length is
a non-universal number the reduced temperature $(T/T_{KT}-1)$ has to
be rescaled by an appropriate factor in order to achieve an acceptable
data collapse. The rescaling factor turns out to be quite large,
namely 18, which is not unusual for microscopically
different systems in the KT universality class (see e.g.,\ 
\cite{pierson}). Note that the finite linear size of the crystallites
plays a similar role as the finite lattice sizes in the simulations
and sets the saturation value for the correlation length (divided by
the lattice constant for the triangular lattice of the CF$_3$Br
molecules, which is $a=4.92\;$\AA). The nice collapse of the experiment
and simulation data supports our claim that the two physical systems are 
in the same universality class and that the orientational ordering of
CF$_3$Br molecules on graphite is in the KT universality class.

\section{Summary}\label{sec:6}

To summarize we have studied the phase transitions in the anti-ferromagnetic
six-state clock model on a triangular lattice, which is fully frustrated. 
As a result the ground states have a $C_6$ (six-state clock) symmetry and a 
$Z_2$ (Ising) symmetry. 
Through extensive Monte Carlo simulations we found that 
the model undergoes a Kosterlitz-Thouless transition at $T_{KT}$ 
and an Ising transition
at $T_I$. The two transitions correspond to the $C_6$ and the $Z_2$ symmetry
breaking transition, respectively. High-precision Monte Carlo data indicate
that the two transitions take place at different temperatures, $T_{KT} <
T_I$~(Eqs.~(\ref{T_I}) and (\ref{T_KT})). 
This has been checked explicitly by analyzing the behavior of the spin 
and the chirality correlation function at temperatures between $T_{KT}$
and $T_I$ (Fig.~\ref{corr_1.037}). 
Furthermore, we have shown that the $Z_2$ symmetry
breaking transition belongs to the Ising universality class. 
For small system sizes, the scaling property of 
the specific heat and the correlation length deviates apparently 
from the Ising universality class.
However, simulation results for larger system sizes indicate that the
model belongs asymptotically to the Ising universality 
class~(Figs.~\ref{cv_star} and \ref{xi_h}).
As for the transition at $T_{KT}$, we have found that the magnetization 
correlation length and the susceptibility diverges at $T=T_{KT}$ according
to the KT scaling form~(Fig.~\ref{KTscaling}). We have also found that
the spins have a quasi long range order below $T_{KT}$~(Fig.~\ref{mA}). 
Combining these,
we conclude that the transition at $T_{KT}$ belongs to the KT universality
class.

Our model is a variant of the fully frustrated $XY$ models where the KT type
ordering and the Ising type ordering interplay interestingly. Our numerical
results support a scenario that there are two separate phase transitions with
$T_{KT}\neq T_I$; one at $T_{KT}$ in the KT universality class and 
the other at $T_I$ in the Ising universality class. 
Our results are consistent with those in Ref.~\cite{Olsson95} very well.

We have proposed that our theoretical model describes the orientational
ordering transition of CF$_3$Br molecules on graphite since the model
has the same symmetry as the experimental system.
We argue that the orientational ordering transition 
belongs to the KT universality class~\cite{Fassbender02} by
comparing the magnetic correlation length obtained from the
experiment~\cite{Fassbender02} with the correlation length 
obtained numerically 
in the six-state clock model. With a suitable rescaling of parameters,
we have shown that the correlation lengths in both systems have the same
scaling behavior~(Fig.\ref{exp-sim}). It gives more evidence that the
orientational ordering transition is indeed the KT transition.


\begin{references}

\bibitem{Teitel83} 
  S. Teitel and C. Jayaprakash, 
  Phys. Rev. B {\bf 27}, 598 (1983).

\bibitem{lee}
  D.H. Lee, J.D. Joannopoulos, J.W. Negele, D.P. Landau, 
  Phys. Rev. Lett. {\bf 52}, 433 (1984);  
  Phys. Rev. B {\bf 33}, 450 (1986).

\bibitem{Choi85&Yosefin85}
  M.Y. Choi and D. Stroud, Phys. Rev. B {\bf 32}, 5773 (1985);
  M. Yosefin and E. Domany, Phys. Rev. B {\bf 32}, 1778 (1985).

\bibitem{Berge86&Eikmans89}
  B. Berge, H.T. Diep, A. Ghazali, and P. Lallemand, Phys. Rev. B {\bf 34},
  3177 (1986); H. Eikmans, J.E. van Himbergen, H.J.F. Knops, and J.M.
  Thijssen, Phys. Rev. B {\bf 39}, 11759 (1989).

\bibitem{JLee91} J. Lee, J. M. Kosterlitz, and E. Granato,
  Phys. Rev. B {\bf 43}, 11531 (1991).

\bibitem{R-Santiago92} G. Ramirez-Santiago and J.V. Jos\'e,
  Phys. Rev. Lett. {\bf 68}, 1224 (1992); Phys. Rev. B {\bf 49}, 9567
  (1994).

\bibitem{Granato93} E. Granato and M.P. Nightingale,
  Phys. Rev. B {\bf 48}, 7438 (1993).

\bibitem{Knops94} Y.M.M. Knops, B. Nienhuis, H.J.F. Knops, and H.W.J. Bl\"ote,
  Phys. Rev. B {\bf 50}, 1061 (1994).

\bibitem{Granato91} E. Granato, J.M. Kosterlitz, J. Lee, and M.P.
  Nightingale, Phys. Rev. Lett. {\bf 66}, 1090 (1991).

\bibitem{Nightingale95} M.P. Nightingale, E. Granato, and J.M. Kosterlitz,
  Phys. Rev. B {\bf 52}, 7402 (1995).

\bibitem{Grest89}
  G.S. Grest, Phys. Rev. B {\bf 39}, 9267 (1989).

\bibitem{JRLee94} 
  J.-R. Lee, Phys. Rev. B {\bf 49}, 3317 (1994).

\bibitem{SLee94} S. Lee and K.-C. Lee,
  Phys. Rev. B {\bf 49}, 15184 (1994).

\bibitem{Miyashita84}
  S. Miyashita and H. Shiba,
  J. Phys. Soc. Jap. {\bf 53}, 1145 (1984).

\bibitem{Xu96}
  H.-J. Xu and B.W. Southern, J. Phys. A {\bf 29}, L133 (1996).

\bibitem{SLee98} S. Lee and K.-C. Lee,
  Phys. Rev. B {\bf 57}, 8472 (1998).

\bibitem{Jeon97} G.S. Jeon, S.Y. Park, and M.Y. Choi,
  Phys. Rev. B {\bf 55}, 14088 (1997).

\bibitem{KT73} J.M. Kosterlitz and D.J. Thouless, 
  J. Phys. C {\bf 6}, 1181 (1973); 
  J.M. Kosterlitz, {\it ibid.} {\bf 7}, 1046 (1974).

\bibitem{Olsson95} P. Olsson,
   Phys. Rev. Lett. {\bf 75}, 2758 (1995).

\bibitem{Comment} J. V. Jos\'e and G. Ramirez-Santiago, Phys. Rev. Lett.
{\bf 77}, 4849 (1995); P. Olsson, Phys. Rev. Lett. {\bf 77}, 4850 (1995).

\bibitem{Boubcheur98} E.H. Boubcheur and H.T. Diep, 
  Phys. Rev. B {\bf 58}, 5163 (1998).

\bibitem{Loison00} D. Loison and P. Simon
  Phys. Rev .B {\bf 61}, 6114 (2000).

\bibitem{Fassbender02}
  S. Fa{\ss}bender, M. Enderle, K. Knorr, J.D. Noh, and H. Rieger,
  Phys. Rev. B, in press (2002).

\bibitem{Jose_etal} 
  J.~V. Jos\'e, L.~P. Kadanoff, S. Kirkpatrick, and D.~R.  Nelson,
  Phys. Rev. B {\bf 16}, 1217 (1977).

\bibitem{Itakura01}
  M. Itakura, J. Phys. Soc. Jpn. {\bf 70}, 600 (2001).

\bibitem{Challa_Landau86} M.S.S. Challa and D.P. Landau,
   Phys. Rev. B {\bf 33}, 437 (1986).

\bibitem{Ferrenberg} A.M. Ferrenberg and R.~H. Swendsen, 
Phys. Rev. Lett.  {\bf 61}, 2635 (1988).


\bibitem{maus}
  E. Maus, Ph.D. thesis, Universit\"at Mainz, 1991.

\bibitem{knorr1}
  K. Knorr, Phys. Rep. {\bf 214}, 113 (1992).

\bibitem{knorr2}
  K. Knorr, S. Fa{\ss}bender, A. Warken, and D. Arndt, 
  J. Low Temp. Phys. {\bf 111}, 339 (1998).

\bibitem{pierson}
  S.W. Pierson, M. Friesen, S.M. Ammirata, J.C. Hunnicutt, and LeRoy A. Gorham, 
  Phys. Rev. B {\bf 60}, 1309 (1999).

\end{references}
\end{document}